\documentclass[runningheads]{llncs}
\usepackage{booktabs, multirow} % for borders and merged ranges
\usepackage{soul}% for underlines
\usepackage[table]{xcolor} % for cell colors
\usepackage{changepage,threeparttable} % for wide tables
\usepackage{graphicx}
\usepackage{caption}
\usepackage{subcaption}
\usepackage{amsmath,amsfonts,bm}

\def\eqref#1{equation~\ref{#1}}
\def\1{\bm{1}}

\def\eps{{\epsilon}}

\def\vx{{\bm{x}}}
\def\vy{{\bm{y}}}

\def\mX{{\bm{X}}}

\DeclareMathAlphabet{\mathsfit}{\encodingdefault}{\sfdefault}{m}{sl}
\SetMathAlphabet{\mathsfit}{bold}{\encodingdefault}{\sfdefault}{bx}{n}

\def\sD{{\mathbb{D}}}
\newcommand{\E}{\mathbb{E}}

\newcommand{\R}{\mathbb{R}}

\newcommand{\KL}{D_{\mathrm{KL}}}

\begin{document}
\title{Cross-adversarial local distribution  regularization for semi-supervised medical image segmentation}
\titlerunning{Cross-adversarial local distribution regularization}
% \author{1030}
% \institute{}
\author{Thanh Nguyen-Duc\inst{1}, Trung Le\inst{1} \and Roland Bammer\inst{1} \and He Zhao\inst{2} \and Jianfei Cai \inst{1} \and Dinh Phung\inst{1}}
\institute{$^1$ Monash University, and $^2$CSIRO's Data61 \\ 
\footnote{$^1$\{\textit{thanh.nguyen4, trunglm, roland.bammer, jianfei.cai, dinh.phung}\}@monash.edu, and $^2$\textit{he.zhao}@ieee.org}}
\authorrunning{T. Nguyen-Duc et al.}

%
%
% \author{Thanh Nguyen-Duc\inst{1} \and
\newcommand{\hz}[1]{\textcolor{red}{HZ:~#1}}

\maketitle              % typeset the header of the contribution
\begin{abstract}
Medical semi-supervised segmentation is a technique where a model is trained to segment objects of interest in medical images with limited annotated data. Existing semi-supervised segmentation methods are usually based on the smoothness assumption. This assumption implies that the model output distributions of two similar data samples are encouraged to be invariant. In other words, the smoothness assumption states that similar samples (e.g., adding small perturbations to an image) should have similar outputs. In this paper, we introduce a novel cross-adversarial local distribution (Cross-ALD) regularization to further enhance the smoothness assumption for semi-supervised medical image segmentation task. We conducted comprehensive experiments that the Cross-ALD archives state-of-the-art  performance against many recent methods on the public LA and ACDC datasets.
%This approach significantly reduces the annotation burden, which is very expensive in the medical domain. 

\end{abstract}
\vspace{-0.6cm}
\section{Introduction}\vspace{-0.2cm}
Medical image segmentation is a critical task in computer-aided diagnosis and treatment planning. It involves the delineation of anatomical structures or pathological regions in medical images, such as magnetic resonance imaging (MRI) or computed tomography (CT) scans. Accurate and efficient segmentation is essential for various medical applications, including tumor detection, surgical planning, and monitoring disease progression. However, manual medical imaging annotation is time-consuming and expensive because it requires the domain knowledge from medical experts. Therefore, there is a growing interest in developing semi-supervised learning that leverages both labeled and unlabeled data to improve the performance of image segmentation models \cite{yang2022survey,ouali2020overview}.

%In existing semi-supervised segmentation methods, smoothness assumption is usually used 
% to enforce the idea
Existing semi-supervised segmentation methods exploit smoothness assumption, e.g.,  the data samples that are closer to each other are more likely to to have the same label.  In other words, the smoothness assumption encourages the model to generate invariant outputs under small perturbations. We have seen such perturbations being be added to natural input images at data-level \cite{miyato2018virtual,french2019semi,li2020transformation,sohn2020fixmatch,wu2022exploring}, feature-level \cite{ouali2020semi,wu2021learning,xie2021intra,lai2021semi}, and model-level \cite{li2020shape,luo2021semi,luo2021efficient,xia20203d,yu2019uncertainty}.  Among them, virtual adversarial training (VAT)~\cite{miyato2018virtual} is a well-known one which promotes the smoothness of the local output distribution using adversarial examples. The adversarial examples are near decision boundaries generated by adding adversarial perturbations to natural inputs. However, VAT can only create one adversarial sample in a run, which is often insufficient to completely explore the space of possible perturbations (see Section \ref{sec:vat}). 
% Even though VAT with random initialization solutions 
In addition, the adversarial examples of VAT can also lie together and lose diversity that significantly reduces the quality of adversarial examples \cite{tashiro2020diversity,nguyen2022particle}.  Mixup regularization \cite{zhang2017mixup} is a data augmentation method used in deep learning to improve model generalization. The idea behind mixup is to create new training examples by linearly interpolating between pairs of existing examples and their corresponding labels, which has been adopted in \cite{berthelot2019mixmatch,berthelot2019remixmatch,sohn2020fixmatch} to semi-supervised learning. The work \cite{gyawali2020enhancing} suggests that Mixup improves the smoothness of the neural function by bounding the Lipschitz constant of the gradient function of the neural networks. However, we show that mixing between more informative samples (e.g., adversarial examples near decision boundaries) can lead to a better performance enhancement compared to mixing natural samples (see Section \ref{sec:ablation}).

In this paper, we propose a novel cross-adversarial local distribution regularization for semi-supervised medical image segmentation for smoothness assumption enhancement \footnote{The Cross-ALD implementation in \textit{https://github.com/PotatoThanh/Cross-adversarial-local-distribution-regularization}}.
Our contributions are summarized as follows: \textbf{1)} To overcome the VAT's drawback, we formulate an adversarial local distribution (ALD) with Dice loss function that covers all possible adversarial examples within a ball constraint.
\textbf{2)} To enhance smoothness assumption, we propose a novel cross-adversarial local distribution regularization (Cross-ALD) to encourage the smoothness assumption, which is a random mixing between two ALDs.
\textbf{3)} We also propose a sufficiently approximation for the Cross-ALD by a multiple particle-based search using semantic feature Stein Variational Gradient Decent (SVGDF), an enhancement of the vanilla SVGD                   \cite{NIPS2016_b3ba8f1b}.
\textbf{4)} We conduct comprehensive experiments on ADCD \cite{bernard2018deep} and LA \cite{xiong2021global} datasets, showing that our Cross-ALD regularization achieves state-of-the-art performance against existing solutions \cite{miyato2018virtual,yu2019uncertainty,li2020shape,luo2021semi,luo2021efficient,wu2021semi,wu2022exploring}.
\vspace{-0.3cm}
\section{Method}\vspace{-0.2cm}
In this section, we begin by reviewing the minimax optimization problem of virtual adversarial training (VAT)\cite{miyato2018virtual}. Given an input, we then formulate a novel adversarial local distribution (ALD) with Dice loss, which benefits the medical semi-supervised image segmentation problem specifically. Next, a cross-adversarial local distribution (Cross-ALD) is constructed by randomly combining two ALDs. We approximate the ALD by a particle-based method named semantic feature Stein Variational Gradient Descent (SVGDF). Considering the resolution of medical images are usually high, we enhance the vanilla SVGD \cite{NIPS2016_b3ba8f1b} from data-level
to feature-level, which is named SVGDF. We finally provide our regularization loss for  semi-supervised medical image segmentation.
\vspace{-0.2cm}
\subsection{The minimax optimization of VAT} \label{sec:vat}\vspace{-0.2cm}
Let $\sD_{l}$ and $\sD_{ul}$ be the labeled and unlabeled dataset, respectively, with $P_{\sD_{l}}$ and $P_{\sD_{ul}}$ being the corresponding data distribution.
Denote $\vx \in \R^d$ as our $d$-dimensional input in a space $\mX$.
% ($\vx \sim P_{\sD}$, where $\sD = \sD_{l} \cup \sD_{ul}$).
The labeled image $\vx_l$ and segmentation ground-truth $\vy$ are sampled from the labeled dataset $\sD_{l}$ ($\vx_l, \vy \sim P_{\sD_{l}}$), and the unlabeled image sampled from $\sD_{ul}$ is $\vx_{} \sim P_{\sD_{ul}}$.

Given an input $\vx \sim P_{\sD_{ul}}$ (i.e., the unlabeled data distribution), let us denote the ball constraint around the image $\vx$ as $C_\eps(\vx) = \{ \vx' \in \mX : || \vx' - \vx ||_p \leq \eps \}$, where $\epsilon$ is a ball constraint radius with respect to a norm $|| \cdot ||_p$, and $\vx'$ is an adversarial example\footnote{A sample generated by adding perturbations toward the adversarial direction.}.
Given that $f_\theta$ is our model parameterized by $\theta$, VAT \cite{miyato2018virtual} trains the model with the loss of $\ell_{vat}$ that a minimax optimization problem:

\begin{equation}\label{eq:vat}
\ell_{vat} := \min_\theta \E_{\vx\sim P_{\sD_{ul}}} \Big[ \max_{\vx' \in C_\eps(\vx)} \KL( f_\theta(\vx'), f_\theta(\vx)) \Big],
\end{equation}
where $\KL$ is the Kullback-Leibler divergence. The inner \textit{maximization problem} is to find an adversarial example near decision boundaries, while the \textit{minimization problem} enforces the local smoothness of the model. However, VAT is insufficient to explore the set of of all adversarial examples within the constraint $C_\eps$ because it only find one adversarial example $\vx'$ given a natural input $\vx$. Moreover, the works \cite{tashiro2020diversity,nguyen2022particle} show that even solving the \textit{maximization problem} with random initialization, its solutions can also lie together and lose diversity, which significantly reduces the quality of adversarial examples.
\vspace{-0.2cm}
\subsection{Adversarial local distribution}\vspace{-0.2cm}
In order to overcome the drawback of VAT, we introduce our proposed adversarial local distribution (ALD) with Dice loss function instead of $\KL$ in \cite{nguyen2022particle,miyato2018virtual}. ALD forms a set of all adversarial examples $\vx'$ within the ball constraint given an input $\vx$. Therefore, the distribution can helps to sufficiently explore all possible adversarial examples.
The adversarial local distribution $P_\theta(\vx'|\vx)$ is defined with a ball constraint $C_\epsilon$ as follow:
\begin{equation}\label{eq:ad}
	P_\theta (\vx' |  \vx) := \frac{e ^ { \ell_{Dice}( \vx' , \vx ; \theta)} }   { \int_{C_\eps(\vx)}    e ^ { \ell_{Dice}( \vx'', \vx ; \theta)} d\vx''  } = \frac{e ^ { \ell_{Dice}( \vx', \vx ; \theta)} }   { Z{(\vx;  \theta)}  },
\end{equation}
where $P_\theta ( \cdot |  \vx)$ is the conditional local distribution, and $Z{(\vx;  \theta)}$ is a  normalization function. The $\ell_{Dice}$ is the Dice loss function as shown in Eq.~\ref{eq:dice}
\begin{equation}\label{eq:dice}
	\ell_{Dice}( \vx' , \vx ; \theta) = \frac{1}{C} \sum_{c=1}^C [ 1 - \frac{ 2 ||p_\theta(\hat{\vy}_c | \vx) \cap p_\theta(\tilde{\vy}_c | \vx') ||} {|| p_\theta(\hat{\vy}_c | \vx) + p_\theta(\tilde{\vy}_c | \vx') ||} ],
\end{equation}
where $C$ is the number of classes. $p_\theta(\hat{\vy}_c | \vx)$ and $p_\theta(\tilde{\vy}_c | \vx')$ are the predictions of input image $\vx$ and adversarial image $\vx'$, respectively.
% \vspace{-0.1cm}
\subsection{Cross-adversarial distribution regularization}\vspace{-0.2cm}
Given two random samples $\vx_i, \vx_j \sim P_{\sD}$  ($i \neq j$), we define the cross-adversarial distribution (Cross-ALD) denoted $\tilde{P}_\theta$ as shown in Eq.~\ref{eq:cross-ad}

\begin{equation}\label{eq:cross-ad}
\tilde{P}_\theta ( \cdot|  \vx_i, \vx_j) = \gamma P_\theta ( \cdot|  \vx_i) + (1-\gamma) P_\theta ( \cdot |  \vx_j)
\end{equation}
where $\gamma \sim$ Beta($\alpha, \alpha$) for $\alpha \in (0, \infty)$, inspired by \cite{zhang2017mixup}. The $\tilde{P}_\theta$ is the Cross-ALD distribution, a mixture between the two adversarial local distributions. \\
Given Eq.~\ref{eq:cross-ad}, we propose the Cross-ALD regularization at two random input images $\vx_i, \vx_j \sim P_{\sD}$ ($i\neq j$) as
\begin{equation}\label{eq:reg}
	\begin{split}
	 R(\theta,\vx_i, \vx_j):= ~ \E_{\tilde{\vx}' \sim  \tilde{P}_\theta (\cdot |  \vx_i,\vx_j)} [  \log \tilde{P}_\theta (\tilde{\vx}' |  \vx_i, \vx_j)   ] ~ =-H(\tilde{P}_\theta (\cdot |  \vx_i, \vx_j)),
	\end{split}
\end{equation}
where H indicates the entropy of a given distribution.

When minimizing $R(\theta,\vx_i, \vx_j)$  or equivalently $ -H(P_\theta (\cdot |  \vx_i, \vx_j))$ w.r.t. $\theta$, we encourage $P_\theta (\cdot |  \vx_i, \vx_j)$ to be closer to a uniform distribution. This implies that the outputs of $f(\tilde{\vx}') = f(\tilde{\vx}'')$ = a constant $c$, where $\tilde{\vx}', \tilde{\vx}'' \sim \tilde{P}_\theta (\cdot |  \vx_i,\vx_j)$. In other words, we encourages the invariant  model outputs under small perturbations.
% This further implies $ \ell( \tilde{\vx}' , \vx_1,\vx_2 ; \theta) = \ell( \tilde{\vx}'' , \vx_1,\vx_2 ; \theta) $ = a constant $c$, where $\tilde{\vx}', \tilde{\vx}'' \sim \tilde{P}_\theta (\cdot |  \vx_1,\vx_2)$.
% In other words, the loss $\ell$ is smooth.
Therefore, minimizing the Cross-ALD regularization loss leads to an enhancement in the model smoothness. While VAT only enforces local smoothness using one adversarial example, Cross-ALD further encourages smoothness of both local and mixed adversarial distributions to improve the model generalization.
\vspace{-0.2cm}
\subsection{Multiple particle-based search to approximate the Cross-ALD regularization}\vspace{-0.2cm}
In Eq.~\ref{eq:ad}, the normalization $Z{(\vx;  \theta)} $ in denominator term is intractable to find. 
% In order to approximate Cross-ALD regularization, we need to find the adversarial local distributions in Eq.~\ref{eq:reg}. However, the $Z{(\vx;  \theta)} $ is intractable to find in Eq.~\ref{eq:ad}. 
Therefore, we propose a multiple  particle-based search method named SVGDF  to sample $\vx'^{(1)}, \vx'^{(2)}, \dots, \vx'^{(N)}  \sim P_\theta (\cdot |  \vx))$. $N$ is the number of samples (or \textit{adversarial particles}).
SVGDF is used to solve the optimization problem of finding a target distribution $ P_\theta (\cdot |  \vx))$.
SVGDF is a particle-based Bayesian inference algorithm that seeks a set of points (or particles)  to approximate the target distribution without explicit parametric assumptions using iterative gradient-based updates.
Specifically, a set of adversarial particles ($\vx'^{(n)}$) is initialized by adding uniform noises, then projected onto the ball $C_\eps$. These adversarial particles are then iteratively updated using a closed-form solution (Eq.\ref{eq:svgd}) until reaching termination conditions (, number of iterations).
\begin{equation}\label{eq:svgd}
\begin{split}
&\vx'^{(n),(l+1)} = \prod_{C_\eps} \Big( \vx'^{(n),(l)} + \tau * \big(\phi (\vx'^{(n),(l)} )  \big) \Big) \\
& s.t. ~ \phi (\vx') = \frac{1}{N} \sum_{j=1}^N  [ k(\Phi(\vx'^{(j),(l) }), \Phi(\vx')) \nabla_{\vx'^{(j),(l)}}  \log P(\vx'^{(j),(l)} | \vx) \\  
&~~~~~~~~~~~~~~~~~~~~ + \nabla_{\vx^{(j),(l)}}  k(\Phi(\vx'^{(j),(l)}), \Phi(\vx')) ],
\end{split}
\end{equation}
where $\vx'^{(n),(l)}$ is a $n^{th}$ adversarial particle at $l^{th}$ iteration ($n \in \{ 1, 2, ..., N\}$, and $l \in \{ 1, 2, ..., L\}$ with the maximum number of iteration $L$).
$\prod_{C_\eps}$ is projection operator to the $C_\eps$ constraint.
$\tau$ is the step size updating.
$k$ is the radial basis function (RBF) kernel $k(\vx', \vx) = \exp \left\{ \frac{-|| \vx' -\vx||^2}{2\sigma^2} \right\}$.
$\Phi$ is a fixed feature extractor (e.g., encoder of U-Net/V-Net). While vanilla SVGD \cite{NIPS2016_b3ba8f1b} is difficult to capture semantic meaning of high-resolution data because of calculating RBF kernel ($k$) directly on the data-level, we use the feature extractor $\Phi$ as a semantic transformation to further enhance the SVGD algorithm performance for medical imaging.
Moreover, the two terms of $\phi$ in Eq.~\ref{eq:svgd} have different roles: (i) the first one encourages the adversarial particles to move towards the high density areas of $P_\theta (\cdot |  \vx)$ and (ii) the second one prevents all the particles from collapsing into the local modes of  $P_\theta (\cdot |  \vx)$ to enhance diversity (e.g.,pushing the particles away from each other).  Please refer to the Cross-ALD Github repository for more details.

SVGDF approximates $P_\theta ( \cdot|  \vx_i)$ and $P_\theta ( \cdot |  \vx_j)$ in Eq.~\ref{eq:cross-ad}, where $\vx_i, \vx_j \sim P_{\sD_{ul}}$ ($i\neq j$). We form sets of adversarial particles as $\sD_{adv}| \vx_i$= \{ $\vx_i'^{(1)}, \vx_i'^{(2)}, \dots, \vx_i'^{(N)}$\} and $\sD_{adv}| \vx_j$ = \{$\vx_j'^{(1)}, \vx_j'^{(2)}, \dots, \vx_j'^{(N)} $\}.
The problem (\ref{eq:reg}) can then be relaxed to 
\begin{equation}\label{eq:reg2}
	\begin{split}
	 R(\theta,\vx_i, \vx_j)&:= \E_{ \vx_i'^{(n)} \sim P_{\sD_{adv}|\vx_i}, \vx_j'^{(m)} \sim P_{\sD_{adv}|\vx_j}} 
  \Big[ \ell_{Dice}(\tilde{x}', \tilde{x}; \theta) \Big] \\
  & s.t.:\tilde{x}' = \gamma \vx_i'^{(n)} + (1-\gamma) \vx_j'^{(m)};~ \tilde{x} = \gamma \vx_i + (1-\gamma) \vx_j,
	\end{split}
\end{equation}
where $\gamma \sim$ Beta($\alpha, \alpha$) for $\alpha \in (0, \infty)$. 
% Note that instead of directly solving (\ref{eq:reg}), we mix adversarial particles sampling from the adversarial local distributions.
\vspace{-0.2cm}
\subsection{Cross-ALD regularization loss in medical semi-supervised image segmentation}\vspace{-0.2cm}
In this paper, the  overall loss function $\ell_{total}$ consists of three loss terms. The first term is the dice loss, where labeled image $\vx_l$ and segmentation ground-truth $\vy$ are sampled from labeled dataset $\sD_{l}$. The second term is a contrastive learning loss  for inter-class separation $\ell_{cs}$ proposed by \cite{wu2022exploring}. The third term is our Cross-ALD regularization, which is an enhancement of $\ell_{vat}$ to significantly improve the model performance. 
\begin{equation}\label{eq:semi}
	\begin{split}
	\ell_{total} := \min_\theta ~ \E_ {(\vx_l, \vy) \sim P_{\sD_{l}}} &\big[ l_{Dice}(\vx_l, \vy; \theta) \big ] + \lambda_{cs} ~\E_ {\vx_l \sim P_{\sD_{l}}, \vx_{} \sim P_{\sD_{ul}}} \big[ \ell_{cs}(\vx_l, \vx) \big] \\
  &+ \lambda_{Cross-ALD} ~\E_ {(\vx_i, \vx_j) \sim P_{\sD_{ul}}} \big[ R(\theta,\vx_i, \vx_j) \big],  
	\end{split}
\end{equation}
where $\lambda_{cs}$ and $\lambda_{Cross-ALD}$ are the corresponding weights to balance the losses. Note that our implementation is replacing $\ell_{vat}$ loss with the proposed Cross-AD regularization in SS-Net code repository\footnote{https://github.com/ycwu1997/SS-Net} \cite{wu2022exploring} to reach the state-of-the-art performance.

\vspace{-0.2cm}
\section{Experiments}
\vspace{-0.2cm}
In this section, we conduct several comprehensive experiments using the ACDC\footnote{https://www.creatis.insa-lyon.fr/Challenge/acdc/databases.html} dataset \cite{bernard2018deep}  and the LA \footnote{ http://atriaseg2018.cardiacatlas.org} dataset \cite{xiong2021global} for 2D and 3D image segmentation tasks, respectively. 
For fair comparisons, all experiments are conducted using the identical setting, following \cite{wu2022exploring}. We evaluate our model in challenging semi-supervised scenarios, where only 5\% and 10\% of the data are labeled and the remaining data in the training set is treated as unlabeled. The Cross-ALD uses the U-Net \cite{ronneberger2015u} and V-Net \cite{milletari2016v} architectures for the ACDC and LA dataset, respectively.
We compare the diversity between the adversarial particles generated by our method against vanilla SVGD and VAT with random initialization in Section \ref{sec:diverse} . We then illustrate the Cross-AD outperforms other recent methods on ACDC and LA datasets in Section \ref{sec:performance}. We show ablation studies in Section \ref{sec:ablation}. The effect of the number particles to the model performance is studied in the Cross-ALD Github repository.  
% Our experiment setups and parameter settings are similar to the work \cite{wu2022exploring} for fair comparisons. 
% Please refer to the supplementary material for all experimental settings.
\vspace{-0.5cm}
\subsection{Diversity of adversarial particle comparison}\label{sec:diverse}
\begin{figure}[!htp]
\vspace{-0.6cm}
 \centering	
 \begin{subfigure}{0.4\linewidth}
 	\centering
 	\includegraphics[width=1\linewidth]{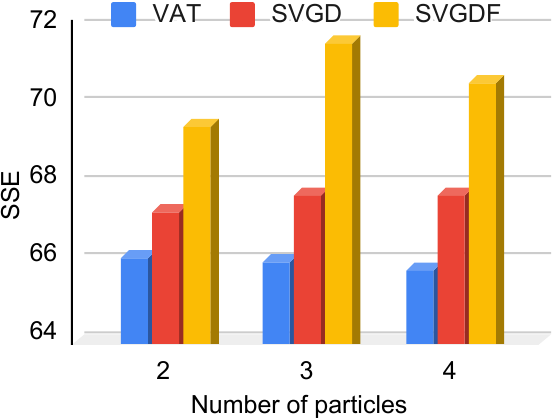}
	\caption{ACDC}
% 	\label{fig:mnist1}
 \end{subfigure}
 \hfill
 \begin{subfigure}{0.4\linewidth}
 	\centering
 	\includegraphics[width=1\linewidth]{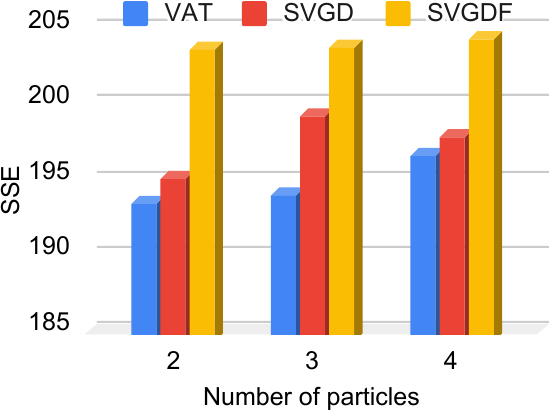}
	\caption{LA}
    % \label{fig:cifar101}
 \end{subfigure}
 \caption{Diversity comparison of our SVGDF, SVGD and VAT with random initialization using sum of square error (SSE) of ACDC and LA datasets.}
 \label{fig:sse}
 \vspace{-0.6cm}
\end{figure}
\textbf{Settings.}
We fixed all the decoder models (U-Net for ACDC and V-Net for LA). We run VAT with random initialization and SVGD multiple times  to produce adversarial examples, which we compared to the adversarial particles generated using SVGDF. SVGDF is the proposed algorithm, which leverages feature transformation to capture the semantic meaning of inputs.  $\Phi$ is the decoder of U-Net in ACDC dataset, while $\Phi$ is the decoder of V-Net  in LA dataset.    We set the same radius ball constraint, updating step, and etc.. We randomly pick three images from the datasets to generate adversarial particles. To evaluate their diversity, we report the sum squared error (SSE) between these particles. Higher SSE indicates more diversity, and for each number of particles, we calculate the average of the mean of SSEs.

\textbf{Results.}
Note that the advantage of SVGD over VAT is that the former generates diversified adversarial examples because of the second term in Eq.~\ref{eq:svgd} while VAT only creates one example. Moreover, vanilla SVGD is difficult to capture semantic meaning of high-resolution medical imaging because it calculates kernel $k$ on image-level. In Fig.~\ref{fig:sse}, our SVGDF produces the most diverse particles compared to SVGD and VAT with random initialization. 
\vspace{-0.5cm}
\subsection{Performance evaluation on the ACDC and LA datasets} \label{sec:performance}
\begin{table}[!t]\centering
\caption{Performance comparisons with six recent methods on ACDC dataset. All results of existing methods are used from \cite{wu2022exploring} for fair comparisons.}\label{tab:ACDC}
\tiny
\begin{tabular}{c|cc|cccc|ccc}\toprule
\multirow{2}{*}{Method} &\multicolumn{2}{c|}{\# Scans used} &\multicolumn{4}{c|}{Metrics} &\multicolumn{2}{c}{Complexity} \\  \cmidrule{2-9}
&Labeled &Unlabeled &Dice(\%)$\uparrow$ &Jaccard(\%)$\uparrow$ &95HD(voxel)$\downarrow$ &ASD(voxel)$\downarrow$ &Para.(M) &MACs(G) \\ \hline
U-Net &3(5\%) &0 &47.83 &37.01 &31.16 &12.62 &1.81 &2.99 \\
U-Net &7(10\%) &0 &79.41 &68.11 &9.35 &2.7 &1.81 &2.99 \\
U-Net &70(All) &0 &91.44 &84.59 &4.3 &0.99 &1.81 &2.99 \\ \hline
UA-MT \cite{yu2019uncertainty}  &\multirow{7}{*}{3 (5\%)} &\multirow{7}{*}{67(95\%)} &46.04 &35.97 &20.08 &7.75 &1.81 &2.99 \\
SASSNet \cite{li2020shape}  & & &57.77 &46.14 &20.05 &6.06 &1.81 &3.02 \\
DTC \cite{luo2021semi}  & & &56.9 &45.67 &23.36 &7.39 &1.81 &3.02 \\
URPC \cite{luo2021efficient} & & &55.87 &44.64 &13.6 &3.74 &1.83 &3.02 \\
MC-Net \cite{wu2021semi}  & & &62.85 &52.29 &7.62 &2.33 &2.58 &5.39 \\
SS-Net \cite{wu2022exploring}  & & &65.82 &55.38 &6.67 &2.28 &1.83 &2.99 \\
Cross-ALD (\textbf{Ours}) & & &\textbf{80.6} &\textbf{69.08} &\textbf{5.96} &\textbf{1.9} &1.83 &2.99 \\ \hline
UA-MT \cite{yu2019uncertainty} &\multirow{7}{*}{7 (10\%)} &\multirow{7}{*}{63(90\%)} &81.65 &70.64 &6.88 &2.02 &1.81 &2.99 \\
SASSNet \cite{li2020shape}  & & &84.5 &74.34 &5.42 &1.86 &1.81 &3.02 \\
DTC \cite{luo2021semi}  & & &84.29 &73.92 &12.81 &4.01 &1.81 &3.02 \\
URPC \cite{luo2021efficient} & & &83.1 &72.41 &4.84 &1.53 &1.83 &3.02 \\
MC-Net \cite{wu2021semi}  & & &86.44 &77.04 &5.5 &1.84 &2.58 &5.39 \\
SS-Net \cite{wu2022exploring}  & & &86.78 &77.67 &6.07 &\textbf{1.4} &1.83 &2.99 \\
Cross-ALD (\textbf{Ours}) & & &\textbf{87.52} &\textbf{78.62} &\textbf{4.81} &1.6 &1.83 &2.99 \\
\bottomrule
\end{tabular}
\vspace{-0.4cm}
\end{table}

\begin{table}[!t]\centering
\caption{Performance comparisons with six recent methods on LA dataset. All results of existing methods are used from \cite{wu2022exploring} for fair comparisons.}\label{tab:LA}
\tiny
\begin{tabular}{c|cc|cccc|ccc}\toprule
\multirow{2}{*}{Method} &\multicolumn{2f}{c|}{\# Scans used} &\multicolumn{4}{c|}{Metrics} &\multicolumn{2}{c}{Complexity} \\  \cmidrule{2-9}
&Labeled &Unlabeled &Dice(\%)$\uparrow$ &Jaccard(\%)$\uparrow$ &95HD(voxel)$\downarrow$ &ASD(voxel)$\downarrow$ &Para.(M) &MACs(G) \\ \hline
V-Net &4(5\%) &0 &52.55 &39.6 &47.05 &9.87 &9.44 &47.02 \\
V-Net &8(10\%) &0 &82.74 &71.72 &13.35 &3.26 &9.44 &47.02 \\
V-Net &80(All) &0 &91.47 &84.36 &5.48 &1.51 &9.44 &47.02 \\ \hline
UA-MT \cite{yu2019uncertainty}  &\multirow{7}{*}{4 (5\%)} &\multirow{7}{*}{76(95\%)} &82.26 &70.98 &13.71 &3.82 &9.44 &47.02 \\
SASSNet \cite{li2020shape}  & & &81.6 &69.63 &16.16 &3.58 &9.44 &47.05 \\
DTC \cite{luo2021semi} & & &81.25 &69.33 &14.9 &3.99 &9.44 &47.05 \\
URPC \cite{luo2021efficient} & & &82.48 &71.35 &14.65 &3.65 &5.88 &69.43 \\
MC-Net \cite{wu2021semi}  & & &83.59 &72.36 &14.07 &2.7 &12.35 &95.15 \\
SS-Net \cite{wu2022exploring} & & &86.33 &76.15 &9.97 &2.31 &9.46 &47.17 \\
Cross-ALD (\textbf{Ours}) & & &\textbf{88.62} &\textbf{79.62} &\textbf{7.098} &\textbf{1.83} &9.46 &47.17 \\ \hline
UA-MT \cite{yu2019uncertainty}  &\multirow{7}{*}{8 (10\%)} &\multirow{7}{*}{72(90\%)} &87.79 &78.39 &8.68 &2.12 &9.44 &47.02 \\
SASSNet \cite{li2020shape}  & & &87.54 &78.05 &9.84 &2.59 &9.44 &47.05 \\
DTC \cite{luo2021semi}  & & &87.51 &78.17 &8.23 &2.36 &9.44 &47.05 \\
URPC \cite{luo2021efficient}  & & &86.92 &77.03 &11.13 &2.28 &5.88 &69.43 \\
MC-Net \cite{wu2021semi}  & & &87.62 &78.25 &10.03 &1.82 &12.35 &95.15 \\
SS-Net \cite{wu2022exploring}  & & &88.55 &79.62 &\textbf{7.49} &1.9 &9.46 &47.17 \\
Cross-ALD (\textbf{Ours}) & & &\textbf{89.92} &\textbf{81.78} &7.65 &\textbf{1.546} &9.46 &47.17 \\
\bottomrule
\end{tabular}
\vspace{-0.6cm}
\end{table}

\textbf{Settings.}
We use the metrics of Dice, Jaccard, 95\% Hausdorff Distance (95HD), and Average Surface Distance (ASD) to evaluate the results. We compare our Cross-ALD to six recent methods including UA-MT \cite{yu2019uncertainty} (MICCAI’19), SASSNet \cite{li2020shape} (MICCAI’20), DTC \cite{luo2021semi} (AAAI’21) , URPC \cite{luo2021efficient} (MICCAI’21) , MC-Net \cite{wu2021semi} (MICCAI’21), and SS-Net \cite{wu2022exploring} (MICCAI'22). 
The loss weights $\lambda_{Cross-ALD}$ and $\lambda_{cs}$ are set as an iteration dependent warming-up function \cite{laine2016temporal}, and number of particles $N$ = 2. 
All experiments are conducted using the identical settings in the Github repository\footnote{https://github.com/ycwu1997/SS-Net} \cite{wu2022exploring} for fair comparisons.\\
\textbf{Results.}
Recall that our Cross-ALD generates diversified adversarial particles using SVGDF compared to vanilla SVGD and VAT, and further enhances smoothness of cross-adversarial local distributions.  
In Table \ref{tab:ACDC} and \ref{tab:LA}, the Cross-ALD can significantly outperform other recent methods with only 5\%/10\% labeled data training based on the four metrics. Especially, our method impressively gains 14.7\% and 2.3\% Dice score higher than state-of-the-art SS-Net using 5\% labeled data of ACDC and LA, respectively.
Moreover, the visualized results of Fig.\ref{fig:visialization} shows Cross-ALD can segment the most organ details compared to other methods.
\vspace{-0.5cm}
\subsection{Ablation study} \label{sec:ablation}
\begin{table}[!htp]\centering
\vspace{-0.6cm}
\caption{Ablation study on ACDC and LA datasets.}\label{tab:ablation}
\tiny
\begin{tabular}{c|c|cc|ccccc}\toprule
\multirow{2}{*}{Dataset} &\multirow{2}{*}{Method} &\multicolumn{2}{c|}{\# Scans used} &\multicolumn{4}{c}{Metrics} \\ \cmidrule{3-8}
& &Labeled &Unlabeled &Dice(\%)$\uparrow$ &Jaccard(\%)$\uparrow$ &95HD(voxel)$\downarrow$ &ASD(voxel)$\downarrow$  \\ \hline
\multirow{7}{*}{ACDC} &U-Net &4(5\%) &0 &47.83 &37.01 &31.16 &12.62 \\\cmidrule{2-8}
&RanMixup &\multirow{6}{*}{4 (5\%)} &\multirow{6}{*}{76(95\%)} &61.78 &51.69 &8.16 &3.44 \\
&VAT & & &63.87 &53.18 &7.61 &3.38 \\
&VAT + Mixup & & &66.23 &56.37 &7.18 &2.53 \\
&SVGD & & &66.53 &58.09 &6.41 &2.4 \\
&SVGDF & & &73.15 &61.71 &6.32 &2.12 \\
&SVGDF + $\ell_{cs}$ & & &74.89 &62.61 &6.52 &2.01 \\
&Cross-ALD (\textbf{Ours}) & & &\textbf{80.6} &\textbf{69.08} &\textbf{5.96} &\textbf{1.9} \\ \hline
\multirow{7}{*}{LA} &V-Net &3(5\%) &0 &52.55 &39.6 &47.05 &9.87 \\ \cmidrule{2-8}
&RanMixup &\multirow{6}{*}{3 (5\%)} &\multirow{6}{*}{67(95\%)} &79.82 &67.44 &16.52 &5.19 \\
&VAT & & &82.27 &70.46 &13.82 &3.48 \\
&VAT + Mixup & & &83.28 &71.77 &12.8 &2.63 \\
&SVGD & & &84.62 &73.6 &11.68 &2.94 \\
&SVGDF & & &86.3 &76.17 &10.01 &2.11 \\
&SVGDF + $\ell_{cs}$ & & &86.55 &76.51 &9.41 &2.24 \\
&Cross-ALD (\textbf{Ours}) & & &\textbf{87.52} &\textbf{78.62} &\textbf{4.81} &\textbf{1.6} \\
\bottomrule
\end{tabular}
\vspace{-0.4cm}
\end{table}

\begin{figure}[!t]
% \vspace{-0.4cm}
 \centering	
 \includegraphics[width=0.9\linewidth]{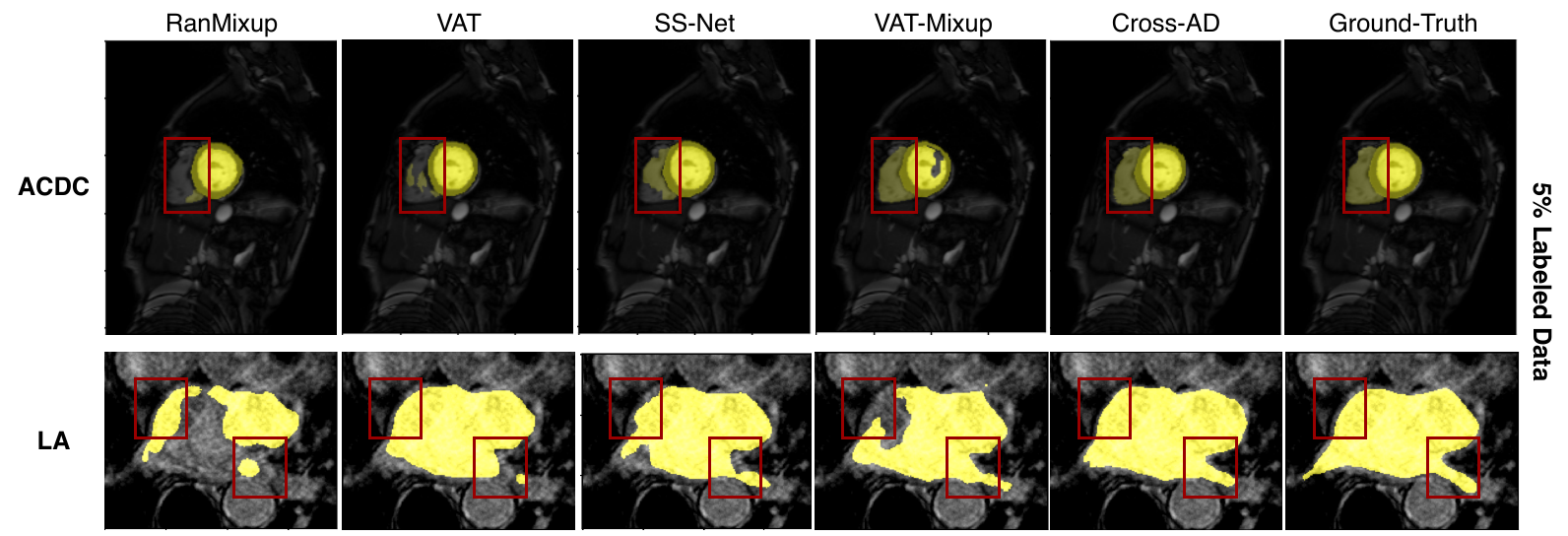}
\caption{Visualization results of several semi-supervised segmentation methods with 5\% labeled training data and its corresponding ground-truth on ACDC and LA datasets.}
\label{fig:visialization}
\vspace{-0.6cm}
\end{figure}
\textbf{Settings.}
We use the same network architectures and parameter settings in Section \ref{sec:performance}, and train the models with 5\% labeled training data of ACDC and LA. We illustrate that crossing adversarial particles is more beneficial than random mixup between natural inputs (RanMixup~\cite{zhang2017mixup}) because these particles are near decision boundaries. Recall that our SVGDF is better than VAT and SVGD by producing more diversified adversarial particles. Applying SVGDF's particles and $\ell_{cs}$ (SVGDF + $\ell_{cs}$ ) to gain the model performance in the semi-supervised segmentation task, while Cross-ALD efficiently enhances smoothness to significantly improve the generalization.  \\
\textbf{Result.} 
Table \ref{tab:ablation} shows that mixing adversarial examples from VAT outperform those from RanMixup. While SVGDF + $\ell_{cs}$  is better than SVGD and VAT,  the proposed Cross-ALD achieves the most outstanding performance among comparisons methods.  In addition, our method produces more accurate segmentation masks compared to the ground-truth, as shown in Fig. \ref{fig:visialization}.

\vspace{-0.5cm}
\section{Conclusion}
\vspace{-0.2cm}
In this paper, we have introduced a novel cross-adversarial local distribution (Cross-ALD) regularization  that extends and overcomes drawbacks of VAT and Mixup  techniques. In our method,  SVGDF is proposed to approximate Cross-ALD, which produces more diverse adversarial particles than vanilla SVGD and VAT with random initialization. We adapt Cross-ALD to semi-supervised medical image segmentation to achieve start-of-the-art performance on the ACDC and LA datasets compared to many recent methods such as VAT \cite{miyato2018virtual},  UA-MT \cite{yu2019uncertainty}, SASSNet \cite{li2020shape}, DTC \cite{luo2021semi}, URPC \cite{luo2021efficient} , MC-Net \cite{wu2021semi}, and SS-Net \cite{wu2022exploring}. 

\subsubsection*{Acknowledgements}
This work was partially supported by the Australian Defence Science and Technology (DST) Group under the Next Generation Technology Fund (NGTF) scheme. Dinh Phung further gratefully acknowledges the partial support from the Australian Research Council, project ARC DP230101176.
% \clearpage
\bibliographystyle{splncs04}
\bibliography{refs}

\end{document}

\title{Cross-adversarial local distribution  regularization for semi-supervised medical image segmentation}
\titlerunning{Cross-adversarial local distribution regularization}
\author{Supplementary}
% \author{Thanh Nguyen-Duc\inst{1} \and
\institute{}
\maketitle              % typeset the header of the contribution
\section{Public ACDC and LA dataset}
\begin{figure}[!htp]
\vspace{-0.4cm}
 \centering	
 \includegraphics[width=1\linewidth]{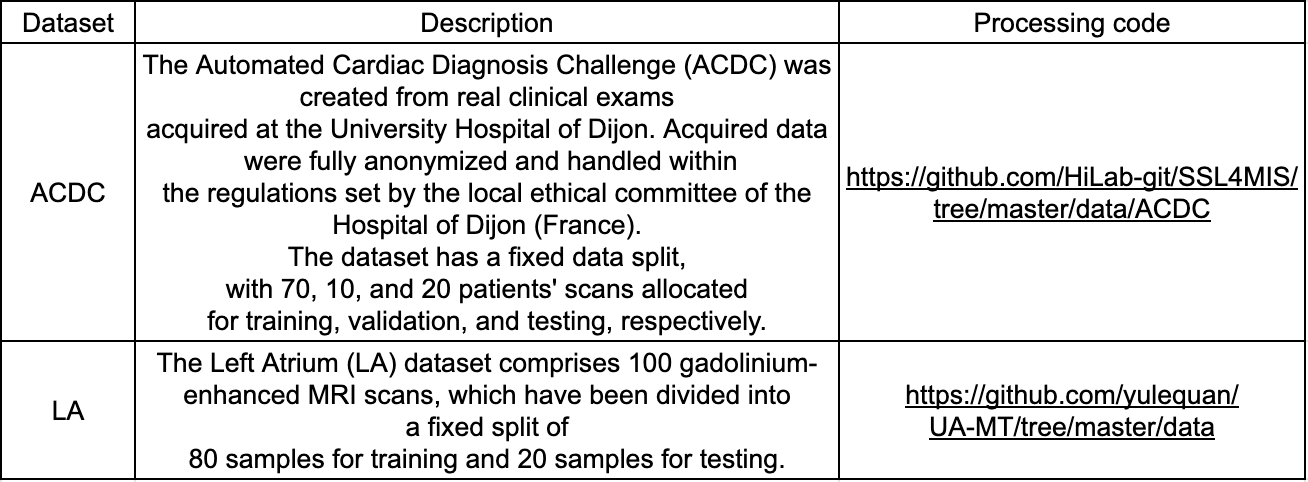}
\caption{Public ACDC and LA datasets.}
\label{fig:visialization}
\vspace{-0.6cm}
\end{figure}

\section{Adversarial particle analysis}\label{sec:analysis}
\begin{table}[!htp]\centering
\caption{We study the number of adversarial particles that affect to the
model performance. We set $N$ in $\{ 1, 2, 3, 4 \}$ for ACDC dataset and $N$ in $\{ 1, 2, 3 \}$ for LA dataset. Note that we use cross-adversarial particles to enhance smoothness. Therefore, by increasing the number of particles, we accordingly increase the regularization strength. The model performance increases by increasing $N$. However,  it is as expected that over regularization may hurt the performance when $N$ = 4 in ACDC dataset.}\label{tab: }
\scriptsize
\begin{tabular}{c|c|cc|ccccc}\toprule
\multirow{2}{*}{Dataset} &\multirow{2}{*}{\# Particles} &\multicolumn{2}{c|}{\# Scans used} &\multicolumn{4}{c}{Metrics} \\ \cmidrule{3-8}
& &Labeled &Unlabeled &Dice(\%)$\uparrow$ &Jaccard(\%)$\uparrow$ &95HD(voxel)$\downarrow$  &ASD(voxel)$\downarrow$  \\ \hline
ACDC &1 &\multirow{4}{*}{3 (5\%)} &\multirow{4}{*}{67(95\%)} &76.59 &65.73 &8.44 &2.21 \\
&2 & & &\textbf{80.6} &\textbf{69.08} &5.96 &\textbf{1.9} \\
&3 & & &80.36 &68.05 &\textbf{5.61} &2.07 \\
&4 & & &77.86 &65.49 &6.16 &2.14 \\ \hline
\multirow{3}{*}{LA} &1 &\multirow{3}{*}{4 (5\%)} &\multirow{3}{*}{76(95\%)} &86.83 &77.03 &5.5671 &1.993 \\
&2 & & &87.52 &\textbf{78.62} &\textbf{4.81} &\textbf{1.6} \\
&3 & & &\textbf{87.71} &78.44 &5.204 &1.9216 \\
\bottomrule
\end{tabular}
\end{table}
% \footnote{We cannot run $N$=4 in 3D segmentation task of LA dataset with the equal batch size due to the limitation of GPU memory.}
% \textbf{Settings.}
% We study the number of adversarial particles that affect to the model performance. The settings are kept similar to Section \ref{sec:performance} except the number of particles $N$. We set $N$ in $\{ 1, 2, 3, 4 \}$ for ACDC dataset and $N$ in $\{ 1, 2, 3 \}$ for LA dataset.

% \textbf{Results.}
% Note that we use cross-adversarial particles to enhance smoothness. Therefore, by increasing the number of particles, we accordingly increase the regularization strength. The model performance increases by increasing $N$. However,  it is as expected that over regularization may hurt the performance when $N$ = 4 in ACDC dataset\footnote{We cannot run $N$=4 in 3D segmentation task of LA dataset with the equal batch size due to the limitation of GPU memory.}.  With more adversarial particles, our method may have increased training time. However, the model inference time remains unchanged  because we do not use these particles during the inference stage.
\clearpage
\section{Semantic feature Stein Variational Gradient Decent (SVDGF)}
\begin{algorithm}[!htp]
    
	\caption{Approximating the adversarial local distribution (ALD) given $\vx$ by using sematic feature Stein Variational Gradient Decent (SVGDF).}\label{alg:svgd}
	\KwIn{A natural sample $\vx \sim P_{\sD_{ul}}$; $n$ number of adversarial particles; $\eps$ for the constraint $C_\eps$; $r$ is $\ell_2$ normalization function;  $\eta$ initial noise factor; $\tau$ step size updating; $L$ number of iterations;  $k$ is RBF kernel function. $\Phi$ is a semactic feature extractor. }
	\KwOut{Set of adversarial particles $ \{\vx'_1, \vx'_2, \dots, \vx'_n \} \sim P_\theta (\cdot |  \vx)$}
	
	Initialise a set of $n$ particles and project to the $B_\eps$ constraint $\{ \vx'_i \in \R^d, i \in \{1, 2, \dots , n\} | \vx'_i = \prod_{C_\eps} (\vx + \eta* Uniform\_noise) \}$\; 
	\For{$l=1$ to $L$}
	{
		\For{ each particle $\vx'^{(l)}_i$}
		{	
			$\vx'^{(l+1)}_i = \prod_{C_\eps} \Big( \vx'^{(l)}_i + \tau * r \big(\phi (\vx'^{(l)}_i )  \big) \Big)$ \;
			where $\phi (\vx') = \frac{1}{n} \sum_{j=1}^n  [ k(\Phi(\vx'^{(l)}_j), \Phi(\vx')) \nabla_{\vx'^{(l)}_j}  \log P(\vx'^{(l)}_j | \vx) +  \nabla_{\vx^{(l)}_j}  k(\Phi(\vx'^{(l)}_j), \Phi(\vx')) ]$ \;
			
		}
	}
	return $ \{\vx'^L_1, \vx'^L_2, \dots, \vx'^L_n \}$ \;

\end{algorithm}